\documentclass[aps,prl,twocolumn,showpacs,groupedaddress,preprintnumbers]{revtex4}

\usepackage{graphicx}

\begin{document}

%
\preprint{MIFP-08-10}

\title{Large Phase of $B_s$-$\bar B_s$ Mixing in Supersymmetric Grand Unified Theories}


\author{Bhaskar Dutta}
\author{Yukihiro Mimura}
\affiliation{
Department of Physics, Texas A\&M University,
College Station, TX 77843-4242, USA
}


\date{\today}

\begin{abstract}

We consider the possibility of a large phase of $B_s$-$\bar B_s$ mixing in supersymmetric SU(5) and SO(10) 
models. We find that in the SU(5) model, the magnitude of this phase is correlated with the branching ratio of $\tau\to\mu\gamma$ and the phase can be within 2$\sigma$ of the recent UTfit analysis. In the case of SO(10) models, this correlation can be relaxed and  a large  phase can be obtained. In this scenario, a non-zero value of  CP asymmetry for $B\to X_s\gamma$ will be predicted. We  predict the sparticle mass ranges for the LHC for these models once the UTfit result is accommodated and discuss the dark matter  and the anomalous magnetic moment constraints on this analysis.

\end{abstract}

\pacs{12.10.Dm, 12.60.Jv, 12.15.Ff}
%

\maketitle


Recently, CDF and D$\O$ collaborations
have announced the analysis
of the flavor-tagged $B_s \to J/\psi \phi$ decay. The decay width difference and the mixing induced
CP violating phase, $\phi_s$, were extracted from their analysis \cite{Aaltonen:2007he}.
In the Standard Model (SM),
the CP violating phase is predicted to be
small,
$\phi_s = 2\beta_s \equiv 2\, {\rm arg}\, (-V_{ts}V_{tb}^*/V_{cs}V_{cb}^*) \simeq 0.04$.
However, the measurements of the phase are large:
\begin{eqnarray}
\phi_s ({\rm CDF}) &\in& [0.32,2.82]\ \ (68\%\, {\rm CL}) \\
\phi_s ({\rm D\O}) &=& 0.57^{+0.30}_{-0.24}({\rm stat})
                       {}^{+0.02}_{-0.07}({\rm syst})
\end{eqnarray}
The UTfit group made a combined data analysis
including the semileptonic asymmetry in the $B_s$ decay,
and find that the CP violating phase deviates more
than $3\sigma$ from the SM prediction \cite{Bona:2008jn}.
This implies the existence of new physics (NP) and
that the NP model requires a flavor violation in $b$-$s$ transition.

Supersymmetry (SUSY) is the most attractive candidate to construct the NP models.
In SUSY models, the flavor universality is often assumed
in squark and slepton mass matrices
to avoid  large flavor changing neutral currents (FCNC)
in the meson mixings and the lepton flavor violations (LFV) \cite{Gabbiani:1988rb}.
Even if we assume the flavor universality,
the non-universality is generated from the evolution of 
renormalization group equations (RGE).
In the minimal extension of SUSY standard model (MSSM)
with right-handed neutrinos,
the induced FCNCs from RGE effects are not large in the quark sector,
while sizable effects can be generated in the lepton sector due to 
the large neutrino mixing angles \cite{Borzumati:1986qx}.
In the grand unified theories (GUT), 
the loop effects due to the large neutrino mixings
can induce sizable effects in the quark sector
since GUT scale particles can propagate in the loops.
The patterns of the induced FCNCs
highly depend on the unification scenario,
and therefore, it is important to investigate
the FCNC effects to obtain a footprint of the GUT models.

If the quark-lepton unification is manifested
in GUT models,
the flavor violation in $b$-$s$ transition
can be responsible for the large atmospheric neutrino mixing \cite{Moroi:2000tk},
and 
thus, it has to be related to the $\tau \to \mu\gamma$ decay \cite{Dutta:2006gq}.
The branching ratio of the $\tau \to \mu\gamma$
is being measured at the $B$-factory,
and thus, the future results of LFV and the 
phase of $B_s$-$\bar B_s$ mixing
will provide an important information to probe
the GUT scale physics.
In this Letter,
we study the correlation between Br($\tau\to\mu\gamma$)
and $\phi_s$
in SU(5) and SO(10) GUT models,
and investigate the constraints in these models from the observations
in order to decipher GUT models.

We first describe the SU(5) and SO(10) GUT models
which we investigate in this Letter.
In the SU(5) model,
the right-handed down-type quarks $(D^c)$
and left-handed lepton doublets ($L$)
are unified in $\bar{\bf 5}$ representation.
The quark doublets $(Q)$, right-handed up-type quarks $(U^c)$,
and right-handed charged-leptons $(E^c)$
are unified in $\bf 10$,
and the right-handed neutrino $(N^c)$ is a singlet under SU(5).
The superpotential which involves the Yukawa interaction is:
$W_Y = Y_u^{ij} {\bf 10}_i {\bf 10}_j H_{\bf 5}
      + Y_d^{ij} {\bf 10}_i \bar {\bf 5}_j H_{\bar {\bf 5}}
      + Y_\nu^{ij} \bar {\bf 5}_i N_j^c H_{\bf 5}$,
where $i,j$ denote the generation indices,
and $H_{\bf 5}$ and $H_{\bar {\bf 5}}$ are the Higgs fields in which colored Higgs fields are unified
with the Higgs doublets.
The charged-lepton Yukawa coupling $Y_e$
is unified to the down-type quark Yukawa coupling $Y_d$, $Y_e = Y_d^{\rm T}$,
in the minimal SU(5) setup.
In the basis where $Y_e$ and Majorana right-handed neutrino mass matrix $M_N$ are diagonal,
the Dirac neutrino Yukawa coupling is denoted as $Y_\nu = U_L Y_\nu^{\rm diag} U_R^{\rm T}$.
When $U_R = {\bf 1}$, the unitary matrix $U_L$ is same as a mixing matrix
of neutrino oscillation,
and it contains large mixing angles.
The Dirac neutrino Yukawa interaction can generate
off-diagonal elements of the (squared) scalar mass matrix for $\bar {\bf 5}$, $M_{\bar {\bf 5}}^2$,
due to a loop diagram in which $N^c$ and $H_{\bf 5}$ propagate.
The induced flavor violating term in the scalar mass matrix is proportional to
$Y_\nu Y_\nu^\dagger$,
and $M_{\bar {\bf5}}^2$ can be parameterized as
$M_{\bar{\bf 5}}^2 = m_5^2 [{\bf 1} - \kappa U_L {\rm diag} (k_1,k_2,1) U_L^{\dagger}]$,
where $m_5$ is a scalar mass for $\bar {\bf 5}$ 
at a cutoff scale, $M_*$, which is possibly the Planck scale,
and
$\kappa$ represents the size of flavor violation in the scalar mass
and $\kappa \simeq (Y_{\nu}^{\rm diag})_{33}^2/(8\pi^2) (3+ A_0^2/m_0^2) \ln M_*/M_{\rm GUT}$
in the universal SUSY breaking scenario where $m_0$ is a universal scalar mass
and $A_0$ is a universal trilinear coupling.
Assuming the Dirac neutrino Yukawa coupling is hierarchical ($k_1, k_2 \ll 1$),
we obtain the 23 element of $M_{\bar {\bf 5}}^2$ as
$-1/2 \,m_5^2\, \kappa \sin 2\theta_{23}\, e^{i\alpha}$,
where $\theta_{23}$ is a 2-3 mixing in $U_L$,
which is large and responsible for the large atmospheric neutrino mixing
unless there exists a fine-tuned relation among $Y_\nu^{\rm diag}$ and $M_N$.
The phase $e^{i \alpha}$ generates a phase of SUSY contribution for
$B_s$-$\bar B_s$ mixing amplitude, $M_{12}^{\rm SUSY}$,
and the absolute value of $M_{12}^{\rm SUSY}$ is controlled by $\kappa \sin 2\theta_{23}$. 
The off-diagonal elements of the scalar mass matrix for $\bf 10$
are also generated by colored Higgs loop,
in addition to the MSSM contribution,
but they are small since they get generated from quark mixings.

In the SO(10) GUT model, which we use in this Letter,
all matter species are unified in the spinor representation $\bf 16$ \cite{Babu:1992ia}.
The matter representation can couple to the $\bf 10$, $\overline{\bf 126}$ and $\bf 120$
Higgs representations,
$W_Y =  h_{ij} {\bf 16}_i {\bf 16}_j {\bf 10}
     +  f_{ij} {\bf 16}_i {\bf 16}_j \overline{\bf 126}
     +  h^\prime_{ij} {\bf 16}_i {\bf 16}_j {\bf 120}$.
The Majorana neutrino mass is generated from the $f$ coupling term
when the $B-L$ direction 
of $\overline{\bf 126}$ gets a vacuum expectation value (VEV).
The fermion Yukawa couplings, $Y_{u,d,e,\nu}$, are given by the linear combinations of
$h$, $f$, $h^\prime$ multiplied by Higgs mixings.
If there is no cancellation among the $h$, $f$, $h^\prime$ couplings,
the Dirac neutrino Yukawa coupling does not have large
mixing in the basis where the charge-lepton mass matrix is diagonal,
due to the presence of right-handed neutrino in ${\bf 16}$.
Thus, the large neutrino mixings may not originate from $Y_\nu$.
However, they can originate from the relative mixing angle of $h$ and $f$
couplings in the model.
Such a structure can be constructed even in the SU(5) models with additional Higgs fields.
The SO(10) model, however, is more predictive~\cite{Dutta:2004wv}
since the Majorana coupling and the contributions to Dirac fermion masses 
are unified to the $f$ coupling. 
In the basis where $Y_e$ is diagonal,
the $f$ coupling is denoted as $U f^{\rm diag} U^{\rm T}$,
and the unitary matrix $U$ is the neutrino mixing matrix when 
the triplet term of the type II seesaw \cite{Schechter:1980gr} is dominant.
Through the loop diagram in which the heavy particles from $\overline{\bf 126}$
(or $\bf 120$) propagate,
flavor violations term can be generated for all sfermion mass matrices
and it is proportional to $f f^\dagger$ (or $h^\prime h^{\prime \dagger}$).
The off-diagonal elements are similarly denoted
 in the SU(5) case.
The phase of SUSY contribution for the $B_s$-$\bar B_s$ mixing amplitude 
$M_{12}^{\rm SUSY}$
can be generated from the phases of 23 elements of $M^2_{\tilde Q}$
and $M_{\tilde D^c}^2$.
The two phases are independent in the SO(10) model in the basis where $Y_d$
is real diagonal,
and the freedom of choosing one of these phases governs the phase of $M_{12}^{\rm SUSY}$.
It is important that
the absolute value of $M_{12}^{\rm SUSY}$
is enhanced
if both left- and right-handed squark mass matrices have
off-diagonal elements \cite{Dutta:2006gq,Dutta:2006zt}.
Therefore, a large phase of $B_s$-$\bar B_s$ mixing
can be expected in SO(10) model
which is much larger compared to the SU(5) model.

Now we discuss the phase of $B_s$-$\bar B_s$ mixing in the GUT models.
We use the model-independent parameterization of the NP contribution:
$C_{B_s} e^{2i\phi_{B_s}} = M_{12}^{\rm full}/M_{12}^{\rm SM}$,
where `full' means the SM plus NP contribution, 
$M_{12}^{\rm full} = M_{12}^{\rm SM}+ M_{12}^{\rm NP}$.
The NP contribution can be parameterized by two real parameters $C_{B_s}$
and $\phi_{B_s}$.
The time dependent CP asymmetry ($S = \sin \phi_s$) in $B_s \to J/\psi \phi$ 
is dictated by the argument of $M_{12}^{\rm full}$ : 
 $\phi_s = - {\rm arg} M_{12}^{\rm full}$,
and thus $\phi_{B_s}$ is the deviation from the SM prediction:
$\phi_s = 2(\beta_s - \phi_{B_s})$.
It is important to note  that the large SUSY contribution is still allowed
even though  the mass difference of  $B_s$-$\bar B_s$ \cite{Abulencia:2006ze}
is fairly consistent with the SM prediction.
This is
because the mass difference can be just  twice the absolute value
of  $M_{12}^{\rm full}$.
The consistency of the mass difference just means $C_{B_s} \sim 1$,
and a large $\phi_{B_s}$ is still allowed.
Actually, when $C_{B_s} \simeq 1$,
the phase $\phi_{B_s}$ is related as
$2\sin \phi_{B_s} \simeq A_{s}^{\rm NP}/A_s^{\rm SM}$,
where $A_s^{\rm NP,SM} = | M_{12}^{\rm NP,SM} |$.
In the model-independent 
global analysis by the UTfit group,
the fit result is
\begin{equation}
A_s^{\rm NP}/A_s^{\rm SM} \in [0.24,1.38] \cup [1.50,2.47]
\end{equation}
at 95\% probability \cite{Bona:2008jn}.
The argument of $M_{12}^{\rm NP}$ being free
in GUT models is due to the phase in off-diagonal elements in 
SUSY breaking mass matrix (in the basis where $Y_d$ is a real diagonal matrix),
and one can choose an appropriate value for the new phase in the NP contribution.
Therefore, the experimental data constrain $A_s^{\rm NP}/A_s^{\rm SM}$,
and that means $\kappa \sin2\theta_{23}$ is constrained for a given SUSY particle spectrum.

Among the flavor violating decay modes, the current bound for Br($\tau \to \mu\gamma$)
is less than $4.5 \times 10^{-8}$ \cite{Hayasaka:2007vc} and 
Br($b \to s\gamma)= (3.55\pm 0.26)\times 10^{-4} $~\cite{Barberio:2007cr}.
In fact, in the models where gaugino masses are unified
at the GUT scale (neglecting anomaly mediated SUSY breaking contribution),
the $\tau \to \mu\gamma$ constraint is stronger than
the $b \to s\gamma$ constraint in most of the parameter space.
This is because the squark masses are raised  much more compared to
the slepton masses due to the gaugino loop contribution
since the gluino is heavier than the Bino and the Wino at low energy.
This gaugino  effect is also important to
allow a large phase in the $B_s$-$\bar B_s$ mixing.
The gaugino loop effects are flavor invisible
and they enhance the diagonal elements of the scalar mass matrices
while keeping the off-diagonal elements unchanged.
If the flavor universal 
scalar masses at the cutoff scale ($m_5$, $m_{10}$ in our notation) become larger,
both Br($\tau\to\mu\gamma)$ and $\phi_{B_s}$ are suppressed.
However, Br($\tau\to\mu\gamma)$ is much more suppressed compared to $\phi_{B_s}$
for a given $\kappa \sin2\theta_{23}$
because the low energy slepton masses are sensitive to  $m_5$ and $m_{10}$
while squark masses are not so sensitive
due to the gluino loop contribution to their masses.

In the SU(5) model,
we find that 
$m_5$ has to be larger than 1.2 TeV in order
to obtain a large $\phi_{B_s}$ at 95\% probability of the UTfit result
if $m_{10}, \mu < 1$ TeV ($\mu$ is the Higgsino mass)
for a universal gaugino mass $m_{1/2} = 300$ GeV
and $\tan\beta = 10$, which is a ratio of Higgs fields' VEVs.
On the other hand, in the SO(10) model,
we find that $m_{16}$ ($= m_5 = m_{10}$)
needs to be larger than 500 GeV if $\mu < 1$ TeV
and $m_{1/2}= 300$ GeV and $\tan\beta =10$.
The reason of a  smaller scalar mass being allowed in the SO(10) model
is due to a left-right enhancement effect in the box diagram.
The constraint on masses based on UTfit result
provides an important guidance to search for the SUSY particles at the LHC.
Actually, in general, if quark-lepton unification is manifested
in the GUT models,
the slepton masses needs to be heavy (especially in the SU(5) model)
in order to suppress $\tau\to\mu\gamma$ and to obtain a large phase of $B_s$-$\bar B_s$ mixing.

The diagrams for $\tau\to\mu\gamma$ which can provide
important effects
are the chargino loop diagrams.
This contribution can be suppressed if $\mu$ and/or $m_5$ are large.
If $m_{10}$ is  small (which means that right-handed sleptons are light),
the Bino diagram can contribute.
In SU(5) model, the off-diagonal elements of the right-handed sleptons are small,
but the Bino diagram can generate LFV through left-right mixing of sleptons.
The amplitude of  $\tau\to\mu\gamma$ is proportional to $\tan\beta$,
while $\phi_{B_s}$ does not depend on $\tan\beta$ much.
Therefore, in order to obtain a larger $\phi_{B_s}$, 
heavier SUSY particles and smaller $\tan\beta$ are favored.
In Fig.1,
we show the contour plot
for $A_s^{\rm NP}/A_s^{\rm SM}$
when the Br($\tau\to\mu\gamma$) bound is saturated.
To draw the figure, 
we choose $m_{1/2} = 300$ GeV, $m_5 = 2$ TeV, and $\tan\beta = 10$.
We assume that SUSY breaking Higgs masses, $m_{H_u}$ and $m_{H_d}$,
are free to make $\mu$ to be a free parameter,
but we assume to $m_{H_u} = m_{H_d}$ at the GUT scale just for simplicity. 
In the figure, the blue shaded region shows $1 \sigma$ allowed region and 
the yellow shaded region is enclosed for $2\sigma$ allowed region.
The right-bottom area in the figure is excluded
since the stability condition of the Higgs potential
is not satisfied.
In the area, closer to the excluded region,
the mass of the charged Higgs boson is light
and therefore, $b\to s\gamma$ becomes large.
If $\mu$ is large, the gluino contribution becomes large
for right-handed operators of $b\to s\gamma$ 
(which is often called $C_{7R}, C_{8R}$)
due to the large left-right mixing for sbottom,
and in this region, $b\to s\gamma$ is more important
rather than $\tau\to\mu\gamma$.
One can see that large $A_s^{\rm NP}/A_s^{\rm SM}> 0.38$ (which is at 68\% probability)
is allowed for a large value of $\mu$.
This is because the  chargino contribution for $\tau\to\mu\gamma$ is suppressed in those area.
The contours are curved down for small $m_{10}$ because 
the Bino diagram starts contributing.
It is worth noting
that the stau coannihilation region (where the lightest stau and the 
lightest neutralino mass difference is $\sim$ 5-15 GeV~\cite{bd}) 
$m_{10} \sim 100$ GeV  to satisfy the dark matter content
is not allowed at 95\% probability in this figure.
To revive the stau coannihilation region, $\tan\beta \sim 5$ is needed
and then scalar trilinear coupling has to be chosen appropriately
to satisfy the lightest Higgs mass bound.
The Higgsino dark matter is not very favored
(though it is allowed for $\tan\beta =10$ in the 95\% probability)
since a small value of $\mu$ does not suppress the chargino contribution for $\tau\to\mu\gamma$.
The stop coannihilation mechanism or $A$-funnel region would be more suitable to  explain the dark matter content
if the quark-lepton unification is manifested and 
the experimental constraint from  flavor violation is considered. A detailed discussion on dark matter constraint will be presented elsewhere.

\begin{figure}[t]
 \center
 \includegraphics[viewport = 25 24 280 220,width=7.0cm]{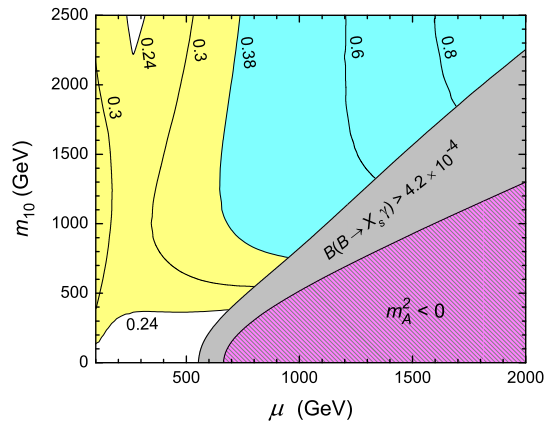}
 \caption{
Contour plot for $A_s^{\rm NP}/A_s^{\rm SM}$ when Br($\tau\to\mu\gamma$)
bound is saturated.
The detail is given in the text.
} 
\end{figure}

Unfortunately, in order to reduce the chargino contribution to $\tau\to\mu\gamma$,
the sleptons must be heavy enough,
and as a result we find that one cannot provide a solution 
for a muon $g-2$ anomaly~\cite{g-2} to obtain large $\phi_{B_s}$
if quark-lepton unification is manifested.
Therefore, we should consider
the possible breaking of the manifest quark-lepton unification.
So, let us consider the possibility that
$\kappa \sin2\theta_{23}$ is different between the squark and slepton 
sectors.
To begin with, the 23 mixing angle $\theta_{23}$
can be different between quarks and leptons
because $Y_d$ and $Y_e$ may not be simultaneously diagonalized.
In minimal SU(5) model, $Y_d$ and $Y_e$ are unified,
but it gives a wrong prediction for the 1st and 2nd generation masses,
and we need a correction from $\bf 45$ Higgs field or non-renormalizable interaction.
Actually, there is a freedom to choose $\theta_{23}^{\rm lepton} \ll \theta_{23}^{\rm quark}$
which is needed to relax the constraint arising from $\tau \to \mu\gamma$.
However, the motivation to explain the large neutrino mixing is lost.
This situation is  same as in the SO(10) models.
Thus, let us consider the possibility that $\kappa$
is different between quark and lepton sectors.
Actually, it should be different
since the flavor violation terms are generated from the loop diagram 
in which heavy particles  propagate,
and the heavy particles should split when GUT symmetry is broken.
In SU(5) model, however, it always gives wrong direction,
i.e., $\kappa^{\rm quark} < \kappa^{\rm lepton}$.
The RGE effect to generate the flavor violation survives 
till the right-handed Majorana mass scale for left-handed slepton,
but it ends at the colored Higgs mass scale for the right-handed down-type squarks.
To satisfy the nucleon decay bounds, the colored Higgs
need to be heavier rather than the right-handed Majorana mass scale
when $Y_\nu$ is less than $O(1)$.
Therefore, in a model, where the flavor violation originates from the 
Dirac neutrino Yukawa coupling,
a large $\phi_{B_s}$ is disfavored.
In this sense, the Fig.1 is drawn conservatively.
When the Majorana mass is $10^{14}$ GeV, the colored Higgs mass is $10^{16}$ GeV,
and the cutoff scale is $10^{18}$ GeV,
one then obtains $2 \kappa^{\rm quark} \simeq \kappa^{\rm lepton}$.
At that time, 
the contour values for $A_s^{\rm NP}/A_s^{\rm SM}$
should be reduced roughly by half  to saturate the $\tau\to\mu\gamma$ bound
and then $\mu$ has to be larger than 1 TeV.

In the SO(10) model, 
$\theta_{23}$'s for quarks and leptons are not  very different
if there is no huge cancellation in the Yukawa couplings, $h$, $f$, and $h^\prime$.
However, 
$\kappa$ can be different among the sfermion species,
and it depends on the SO(10) breaking vacua and the Higgs spectrum from
$\overline{\bf 126}$.
If  the mass of one of the Higgs representation is light
compared to the SO(10) breaking scale,
the off-diagonal elements for some fermion species are generated.
For example,
when the SU(2)$_R$ symmetry remains unbroken and $({\bf 1},{\bf 1},\pm 1)$ fields from $\overline{\bf 126}$,
which are SU(2)$_R$ Higgsinos,
remain light, the off-diagonal elements for only right-handed slepton are generated.
Obviously, these do not generate a large $\phi_{B_s}$.
When $({\bf 8},{\bf 2},\pm 1/2)$ fields of $\overline{\bf 126}$ are light,
only  off-diagonal elements for squarks are generated
while the same  elements for sleptons are small.
This is the proper way to generate large $\phi_{B_s}$
without suffering from the $\tau\to \mu\gamma$ constraint.
It is interesting to note that the light $({\bf 8},{\bf 2},\pm 1/2)$ fields are good candidate 
to suppress the nucleon decay \cite{Dutta:2007ai} in these models.
In this way,
it is possible that 
the experimental measurements
of large $\phi_{B_s}$
and Br($\tau\to\mu\gamma$)
for a given SUSY particle spectrum
can constrain the GUT scale particle spectrum.
Thus, more experimental data is very important to
probe the GUT scale physics.

If a suitable SO(10) breaking vacuum is selected,
the $\tau\to\mu\gamma$ and $\phi_{B_s}$ relation can be completely broken.
If both left- and right-handed squarks have off-diagonal elements
in this vacuum,
the large $\phi_{B_s}$ can be easily obtained.
In Fig. 2,
we demonstrate the allowed region
for different $\kappa m_0^2$.
We assume  a universal scalar mass $m_0 = m_5 = m_{10} = m_{H_u} = m_{H_d}$.
The figure is drawn in the case of $\tan\beta = 10$,
though $\tan\beta$ is not an important parameter in the plot.
When $m_0$ is small, the trilinear scalar coupling $A_0$ has to be large
in order to make $\kappa$ large.
\begin{figure}[t]
 \center
 \includegraphics[viewport = 25 24 280 220,width=7.0cm]{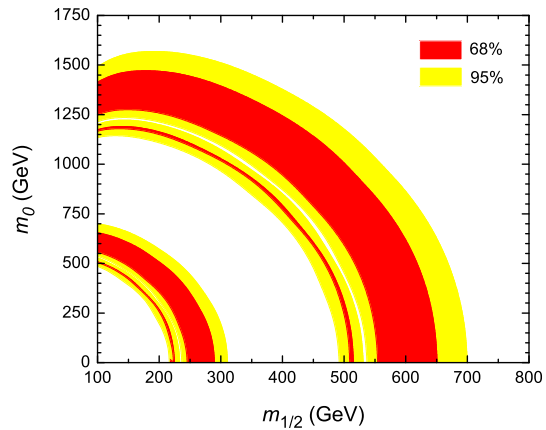}
 \caption{
Allowed region for $\phi_{B_s}$ for $\kappa m_0^2 = (200\ {\rm GeV})^2$ (small quarter circle),
and $\kappa m_0^2 = (600\ {\rm GeV})^2$ (large quarter circle).
} 
\end{figure}
\begin{figure}[t]
 \center
 \includegraphics[viewport = 25 24 280 220,width=7.0cm]{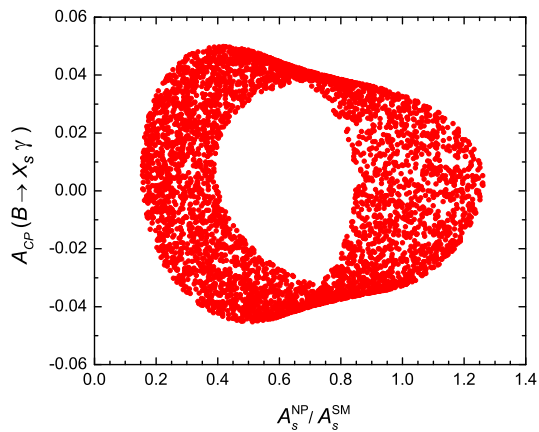}
 \caption{
Direct CP asymmetry of $B \to X_s \gamma$ decay vs
$A_s^{\rm NP}/A_s^{\rm SM}$.
We choose $m_{1/2} = 200$ GeV, $m_0 = 250$ GeV, and $\tan\beta = 40$.
} 
\end{figure}
If  $\tau\to\mu\gamma$ is suppressed by the choice of a vacuum,
the $b \to s\gamma$ constraint becomes more important
especially for large $\tan\beta$.
However, 
since the phase of the chargino contribution is free in the SO(10) boundary condition
(actually it is independent of the phase of $M_{12}^{\rm SUSY}$),
experimentally allowed solutions can be found
as long as the gluino contribution for $C_{7R}$, $C_{8R}$
is not very large.
When the chargino contribution is large,
the direct CP asymmetry for $B \to X_s \gamma$ \cite{Aubert:2004hq} may become large.
This is similar to the case where $\phi_{B_s}$ is large
while $C_{B_s} \sim 1$.
In Fig. 3, we plot the direct CP asymmetry and $A_s^{\rm NP}/A_s^{\rm SM}$
for allowed Br($B\to X_s \gamma$).
Due to the  freedom of the off-diagonal elements of up-type squark
mass matrices,
the $b\to s\gamma$ constraint can be satisfied
even in the usually excluded region in the case of minimal supergravity model.
Actually the parameter $m_{1/2} = 200$ GeV we choose to draw the Fig. 3 
is not allowed in the minimal supergravity for $\tan\beta =40$.
Thus in this region, one needs off-diagonal elements of squark 
mass matrix,
and thus, a  non-zero value of $A_s^{\rm NP}$ is predicted.
If Br($B\to X_s \gamma)$ is fixed to a particular value, 
this plot is just a circle. The dark matter and the anomalous magnetic moment constraints are easily satisfied in this scenario.

In this Letter, we have emphasized the importance of $\tau\to \mu\gamma$
and $\phi_{B_s}$ correlation in GUT models,
since they can be correlated directly by 23 mixing.
The constraint from $\mu\to e\gamma$ may be also important,
but this Br calculation highly depends on the details of flavor structure which can have a freedom of cancellation.
Therefore, we have not talked about the $\mu\to e\gamma$ constraint.
We refer to the Ref.\cite{Dutta:2006zt} for an  analysis of  flavor violation including the first generation.

In conclusion, we have investigated the large phase of $B_s$-$\bar B_s$ mixing by
comparing SU(5) and SO(10) GUT models.
%
The existence of   $\phi_{B_s}$ in GUT models
can tell us  whether the flavor violation 
originates from Dirac neutrino Yukawa coupling or Majorana coupling
such as the $\overline{\bf 126}$ Higgs coupling in the SO(10) model.
At present, the SO(10) models are  more preferred rather than the
Dirac neutrino induced flavor violation in SU(5) models.
It is important to note
that
we can distinguish these two scenarios
once more experimental data on  $B_s$-$\bar B_s$ mixing phase 
and Br($\tau\to\mu\gamma$) decay are available along with the 
data from the LHC.

This work 
is supported in part by the DOE grant
DE-FG02-95ER40917.

\end{document}